%% file: main.tex
\documentclass[runningheads]{llncs}
\usepackage[T1]{fontenc}
\usepackage[numbers]{natbib}
\usepackage{graphicx}
\usepackage{hyperref}
\usepackage{url}
\usepackage{color}

\input{packages}

\begin{document}
\title{WiDiff: Extracting Changes from Wikidata's Edit History}
%
%
\author{Carolina Cortés\inst{1}\orcidID{0009-0002-9539-4796} \and
Lisa Ehrlinger\inst{1}\orcidID{0000-0001-5313-0368} \and
Lorena Etcheverry\inst{2}\orcidID{0000-0001-8121-8076} \and Felix Naumann\inst{1}\orcidID{0000-0002-4483-1389}}
\authorrunning{C. Cortés et al.}
%
\institute{Hasso Plattner Institute, Potsdam, Germany \and
Facultad de Ingeniería, Montevideo, Uruguay
}
\maketitle              
\begin{abstract}
\input{parts/0_abstract}

\keywords{Change \and Evolution \and Knowledge Graphs \and Wikidata}
\end{abstract}

\input{parts/1_introduction}

\input{parts/3_wikidata}

\input{parts/4_preliminaries}

\input{parts/5_change_extraction}

\input{parts/6_change_data}
\input{parts/7_change_analysis}
\input{parts/8_conclusion}
\bibliographystyle{splncs04}
\bibliography{bibliography}
\end{document}

%% file: packages.tex
\usepackage{tikz}
\usepackage{forest}
\usepackage{xcolor}
\usepackage{tabularx}
\usepackage{xcolor}
\usepackage{multirow}
\usepackage{tabularx}
\usepackage{todonotes}
\usepackage{booktabs}
\usepackage{placeins}
\usepackage{float}
\usepackage{xurl}
\usepackage[T1]{fontenc}
\usepackage{numprint}
\npthousandsep{\,}
\usepackage{threeparttable}
\usepackage{url}
\usepackage{makecell}
\usepackage{listing}

\usepackage{xcolor}
\usepackage{listings}

\usepackage{xspace}
\usepackage{algorithm}
\usepackage{algpseudocode}

\newcommand{\schema}[1]{\textsf{\small #1}}

\newcommand{\entity}[1]{\textsf{\small #1}}

\newcommand{\ExternalLink}{%
    \tikz[x=1.2ex, y=1.2ex, baseline=-0.05ex]{%
        \begin{scope}[x=1ex, y=1ex]
            \clip (-0.1,-0.1) 
                --++ (-0, 1.2) 
                --++ (0.6, 0) 
                --++ (0, -0.6) 
                --++ (0.6, 0) 
                --++ (0, -1);
            \path[draw, 
                line width = 0.5, 
                rounded corners=0.5] 
                (0,0) rectangle (1,1);
        \end{scope}
        \path[draw, line width = 0.5] (0.5, 0.5) 
            -- (1, 1);
        \path[draw, line width = 0.5] (0.6, 1) 
            -- (1, 1) -- (1, 0.6);
        }
    }
\newcommand{\wdlink}[1]{\href{#1}{\textcolor{blue}{\ExternalLink}}}

\newcommand{\feature}[1]{\textit{\nolinkurl{#1}}}

\newcommand{\toolname}{WiDiff\xspace}

\usepackage{textcomp}
\usepackage{color}

\definecolor{codegreen}{rgb}{0,0.6,0}
\definecolor{codegray}{rgb}{0.5,0.5,0.5}
\definecolor{codepurple}{HTML}{C42043}
\definecolor{backcolour}{HTML}{F2F2F2}
\definecolor{bookColor}{cmyk}{0,0,0,0.90}  
\color{bookColor}

\lstdefinestyle{sqlstyle}{
    backgroundcolor=\color{backcolour},   
    commentstyle=\color{codegreen},
    keywordstyle=\color{codepurple},
    stringstyle=\color{codepurple},
    basicstyle=\footnotesize\ttfamily,
    breakatwhitespace=false,
    breaklines=true,
    captionpos=b,
    keepspaces=true,
    showspaces=false,
    showstringspaces=false,
    showtabs=false,
}

\newcommand{\mypara}[1]{\smallskip\noindent\textbf{#1.}}
\newcommand{\kgs}{\textsc{kg}s\xspace}
\newcommand{\KG}{\textsc{kg}\xspace}

%% file: parts/0_abstract.tex
Knowledge graphs have become a key resource for integrating heterogeneous data and powering downstream tasks such as question answering, entity linking, and semantic search. 
They are built and maintained incrementally, either (i) fully automated, e.g., YAGO, (ii) semi-automatically with community oversight, e.g., DBpedia, or (iii) manually through collaborative editing, e.g., Wikidata. 
Understanding the \emph{evolution} of knowledge graphs is essential as changes may reflect real-world updates, error corrections, or noise introduced by vandalism, all of which affect the reliability of downstream applications. 
Among openly available knowledge graphs, Wikidata is the most challenging case to study evolution, with over 120 million entities edited by humans and bots and an edit history spanning more than a decade. 
Although Wikidata exposes change data in various formats (e.g., periodic dumps and real-time event streams), none support analytical queries over the complete edit history. 
Therefore, we present \toolname, a tool that extracts changes from Wikidata's complete edit history and provides a unified interface for large-scale analytical queries over it. 

%% file: parts/1_introduction.tex
\section{Introduction}\label{sec:introduction}
Knowledge Graphs (\kgs) have become the foundation for a range of downstream tasks, such as information retrieval and question answering~\cite{Suchanek2024}. 
Real-world \kgs are constantly changing on both data and schema levels, e.g., facts are updated or properties are created. 
These changes are driven by different motivations, such as reflecting real-world events (adding the results of a recent election) or correcting data errors. 
Left unmanaged, such changes can introduce inconsistencies and contradictions that hinder the analysis and reuse of \kgs~\cite{Farber_2018,Piscopo_2019}. 

\mypara{Wikidata}
Wikidata is a collaborative \KG launched by the Wikimedia Foundation in 2012~\cite{VrandecicPK23}. As of August 2025, it represents more than 120~million items, more than 2~billion edits have been made since its start\footnote{\url{https://www.wikidata.org/wiki/Wikidata:Statistics}}, and it continuously receives between 300 and \numprint{1100} edits per 
minute\footnote{\url{https://wikitech.wikimedia.org/wiki/WMDE/Wikidata/Growth\#Edit_rate}} from both humans and bots. Given this collaborative nature, the edits made to Wikidata are diverse in nature and motivation: some edits reflect real-world changes, others are made to improve data quality, or they may also be caused by vandalism and editorial disagreement over how facts should be represented.
Although Wikidata provides access to change data in various formats (e.g., periodic dumps\footnote{\url{https://www.wikidata.org/wiki/Wikidata:Database_download}} and real-time event streams\footnote{\url{https://www.mediawiki.org/wiki/EventStreams}}), none of these formats supports analytical queries over the complete edit history.

\mypara{Related work and research gap}
The study of evolving KGs has been identified as an open problem by Polleres et al.~\cite{pollerespernisch2023}, who discuss the different dimensions of \KG evolution, metrics to measure evolution over time and tools and techniques to store and process evolving \kgs. 
In particular, they identify the need for storage solutions for dynamic and versioned graphs to enable different kinds of analysis as future work.
Bleifuß et al.~\cite{belifuss2018} propose a formal model for storing and querying database changes and a tool that implements them~\cite{bleifuss2019dbchex}, which has been evaluated on several datasets, such as Wikipedia infoboxes and DBLP and serves as basis for our work. 
Käfer et al.~\cite{KaferUHP12} introduced the Dynamic Linked Data Observatory (DyLDO) in 2012, a framework with the original idea to weekly monitor a fixed set of linked data documents from different domains (e.g., \href{http://dbpedia.org}{dbpedia.org}, \url{freebase.com}, or \url{dbtune.org}; but not yet Wikidata).
In the meantime, DyLDO is deprecated and no longer available\footnote{http://swse.deri.org/dyldo/}.
Schmelzeisen et al.~\cite{Wikidated2021} presented with Wikidated 1.0, a dataset of Wikidata's edit history that stores changes between revisions of an entity as sets of triple deletions and additions in  Resource Description Framework (RDF), spanning all revisions from 2012 to 2020. 
However, the dataset has not been publicly released despite being promised on the GitHub repository\footnote{https://github.com/lschmelzeisen/wikidated}, and the code has not been updated since 2022.
Finally, Tanon et al.~\cite{Tanon2019} proposed a system to index Wikidata revisions, enabling the querying of Wikidata's edit history through a SPARQL endpoint. For this, data is stored in the key-value store RocksDB~\footnote{{https://rocksdb.org/}}, and indexes are created on top of this database to enable efficient SPARQL querying. Moreover, due to storage space constraints, their system was demonstrated only for direct claim relations. The endpoint is no longer available\footnote{\url{https://wdhqs.wmflabs.org/}}.

In summary, none of these works \cite{KaferUHP12,Wikidated2021,Tanon2019} provides an up-to-date and maintained tool or dataset that supports large-scale queries over Wikidata's entire edit history.

\mypara{Contribution}
We address this gap with \toolname, an open-source tool\footnote{Code available at \url{https://github.com/caroocortes/WiDiff}} that extracts the complete edit history of Wikidata and provides a unified interface for large-scale analytical queries.
The changed data extracted with \toolname is publicly available at \cite{full_data}.

\mypara{Use cases}
\toolname provides a pipeline for automatically extracting changes from Wikidata, enabling several use cases. For instance, the edit history can be leveraged for \textit{data quality assessment} by tracking how quality has evolved over time, helping identify properties or entity types with persistent quality issues and pinpointing areas where editorial intervention is most needed. Furthermore, the edit history can be used for \textit{knowledge graph evolution}, for instance, to study how coverage across different domains (e.g., scientific articles, cultural heritage) has grown over time, helping identify underrepresented or poorly maintained communities. Finally, the edit history can be leveraged for \textit{edit suggestions}, building on similar approaches developed for Wikipedia tables~\cite{schema_recommendation}, where recommendation rules can be derived from the edit history.

\mypara{Outline}
\autoref{sec:wikidata} introduces the data structure of Wikidata for our subsequent work, while \autoref{sec:preliminaries} introduces terminology used throughout the paper. \autoref{sec:change-model} describes our change data model and \autoref{sec:change_extraction} describes the data extraction process carried out by \toolname.
Finally, \autoref{sec:change_data} presents the dataset obtained with \toolname along with processing information, and \autoref{sec:analysis} presents insights into Wikidata's change activity. \autoref{sec:conclusion} concludes the paper with an outlook on future work.

%% file: parts/3_wikidata.tex
\section{Background on Wikidata's Data Structure}\label{sec:wikidata}
Wikidata organizes its data into pages that correspond to a specific entity, which can represent items (identified by QIDs) or properties (identified by PIDs), and contain statements describing those entities. 
Wikidata uses RDF to model its data; therefore, items can be individuals or classes, and statements are triples composed of a subject, a predicate, and an object. 

Moreover, statements can be annotated with additional information, such as references and qualifiers. 
References point to a source that supports a statement and are modeled as a set of property-value pairs.
Qualifiers provide context for a statement (e.g., time that constraints the temporal validity of a statement) and are modeled as a single property-value pair.

Statements are accompanied by a rank: \textit{normal}, \textit{preferred}, or \textit{deprecated}. 
This rank allows Wikidata to reflect not only the current state of the real world, but also its history, by preserving statements that were once valid but are not anymore, or by marking the preferred value among multiple statements for the same property.

Wikidata defines 18 data types~\cite{help_datatypes}, some of which can have added ``metadata'' (e.g., a value of data type \textit{quantity} is accompanied by a \textit{unit}, \textit{lower} and \textit{upper bound}). In this work, we group Wikidata's data types by their ``JSON type'' as defined in \cite{help_datatypes}. For example, Wikidata's \textit{quantity} data type maps directly to a ``JSON type'' \textit{quantity}, while \textit{geo-shape} has a ``JSON type'' \textit{string}. Therefore, we end up with the following data types: \textit{string}, \textit{quantity}, \textit{time}, \textit{entity}, \textit{globecoordinate}.

%% file: parts/4_preliminaries.tex
\section{Preliminaries}\label{sec:preliminaries}
This section defines the terminology used throughout this paper -- specifically, the notions of edit, revision, diff, and change.
\begin{itemize}
    \item \textit{Edit:} refers to the concrete action any editor can perform on a Wikidata entity (e.g., adding a statement).
    \item \textit{Revision:} A timestamped version of a Wikidata entity, created each time a user submits an edit.
    \item \textit{Diff:} the computed difference between two temporally consecutive revisions of the same entity - the set of additions, deletions, and updates needed to transform one revision into the next.
    \item \textit{Change:} refers to the event of a statement value, rank, qualifier statement value, or reference statement value becoming different between two points in time.
\end{itemize}

\section{Change Data Model}
\label{sec:change-model}
Wikidata provides data dumps in several formats (i.e., JSON, RDF, and XML\footnote{\url{https://www.wikidata.org/wiki/Wikidata:Database_download}}); however, only the XML dumps provide the complete edit history of all pages in Wikidata, whereas JSON and RDF dumps capture only the latest state of pages. 
The edit history dumps consist of several XML files, each containing all revisions for a set of pages (i.e., entity, property, or user talk pages). For entity pages, each revision stores metadata (timestamp, user, comment) and the snapshot of the entity's state at that point in time as a JSON blob following the schema described in~\cite{wikibase_json}. Processing these files means iterating over all pages, and for each page, over all its revisions; therefore, during processing, we can only see the state of a single entity at a specific point in time.

Apart from Wikidata's dump format (XML with JSON embedded), the volume (more than 120 million entities~\cite{wikidata_statistics}) and edit frequency (more than 300 edits per minute~\cite{wikidata_edit_rate}) of Wikidata make storing full graph snapshots impractical, not to mention that each graph snapshot would have to be reconstructed, since this is not provided in the edit history dumps. We therefore chose to store only the changes to an entity's statements, converting our storage target into a change log.

\begin{figure}
    \centering
    \includegraphics[width=\linewidth]{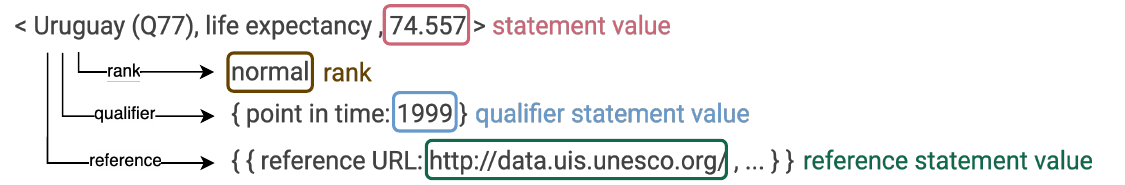}
    \caption{Change granularities captured with WiDiff.}
    \label{fig:change-granularity}
\end{figure}

To represent changes, we adopt the change cube~\cite{change_cube_def}, a generic model for representing changes to a dataset, which uses the tuple $\langle$timestamp, entity\_id, property, new\_value$\rangle$ to represent a change to an entity's property at a point in time. We extend this tuple to represent changes to an entity's statement at different granularities (value, rank, qualifier statement value, and reference statement value -- See \autoref{fig:change-granularity}) by adding (i)~revision metadata (revision identifier, editor, timestamp, and comment), (ii)~the previous value of the statement value, rank, qualifier statement value, or reference statement value, (iii)~data types associated to the old and new values, (iv)~the type of operation (CREATE, DELETE or UPDATE), and (iii)~extra fields that are needed to identify the different granularities. Note that we track changes to a statement's, qualifier's, or reference's \emph{value}: a change to its \emph{property} instead is represented as the deletion of the original and the creation of a new one.


Changes to statement's values or rank are uniquely identified by the tuple $\langle${\small\textit{revision\_id}, \textit{property\_id}, \textit{value\_id}}$\rangle$, where {\small \textit{revision\_id}} identifies the revision, {\small \textit{property\_id}} is the PID for the property, {\small \textit{value\_id}} identifies a specific statement value. 

Unlike statement values, qualifier and reference values in Wikidata are not uniquely identified across revisions. 
Therefore, updates to qualifiers or references cannot be tracked across revisions, since identical values may appear multiple times, and it is not possible to determine which of the values changed without a stable identifier. Therefore, for qualifiers and references, we only capture CREATE and DELETE operations. 
To avoid storing duplicate values, we hash each qualifier or reference value (excluding their metadata); two values are considered distinct only if their hashes differ. 

For qualifier changes, we extend the change tuple with the qualifier's property ID and the qualifier value's hash. Therefore, a change to a qualifier statement value is identified by the tuple $\langle${\small\textit{revision\_id}, \textit{property\_id}, \textit{value\_id}, \textit{qual\_property\_id}, \textit{value\_hash}}$\rangle$.

Conversely, since references in Wikidata consist of sets of statements and we track changes to individual statements within a reference, we require identifiers for the statement value within a reference and for the entire reference. Therefore, for changes to a reference statement value, we extend the change tuple with three values: the reference identifier (the hash of the entire reference), the reference statement value identifier (the hash of the individual statement value), and the reference property ID. Therefore, a change to a reference statement value is identified by the tuple $\langle${\small\textit{revision\_id}, \textit{property\_id}, \textit{value\_id}, \textit{ref\_property\_id}, \textit{ref\_hash}, \textit{value\_hash}}$\rangle$.

%% file: parts/5_change_extraction.tex
\section{\toolname}
\label{sec:change_extraction}

\begin{figure}[h]
    \centering
    \includegraphics[width=\linewidth]{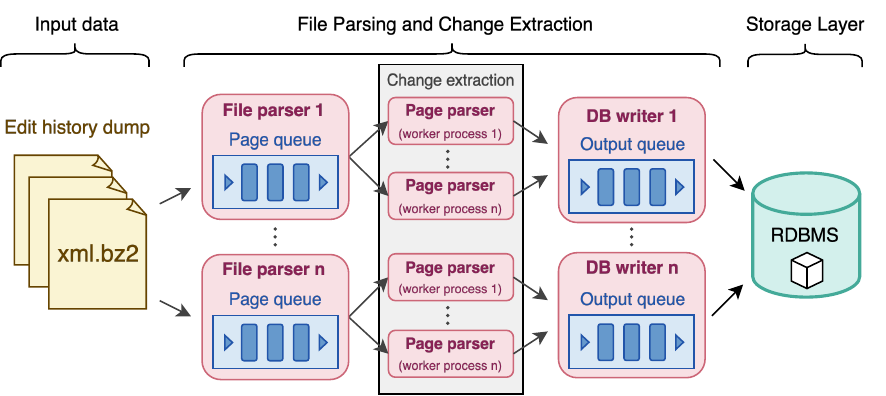}
    \caption{\toolname~architecture.}
    \label{fig:parser_arch}
\end{figure}

\autoref{fig:parser_arch} shows the architecture of \toolname -- each step is described in sections \ref{sec:tool-input} to \ref{sec:change_data}: the input data in \autoref{sec:tool-input}, file parsing and change extraction in \autoref{sec:tool-file-parsing} and \autoref{sec:tool-change-extraction}, the storage layer in \autoref{sec:tool-storage-layer}, and the configuration used to run \toolname in \autoref{sec:tool-config}.

\subsection{Input Data}
\label{sec:tool-input} 
\toolname{} takes as input the \textit{pages-meta-history} files from a Wikidata XML dump, in the .bz2 format. At the time of writing this paper, there were 2125 such files.
Each file contains all revisions for a set of pages, structured as shown in \autoref{lst:xml_structure}. 

\begin{listing}
\caption{Example of XML file structure for the entity Uruguay (Q77).}
\centering
\label{lst:xml_structure}
\begin{lstlisting}[basicstyle=\ttfamily\scriptsize, language=XML]
<page>
    <title>Q77</title>
    <id>77</id>
    <revision>
      <id>1</id>
      <timestamp>2012-10-29T21:30:00Z</timestamp>
      <contributor>
        <username>ExampleUser</username>
        <id>1</id>
      </contributor>
      <comment>Created a country in South America</comment>
      <text xml:space="preserve">
        {"id": "Q77", "labels": {...}, "claims": [], ...}
      </text>
    </revision>
    <revision>...</revision>
    ...
</page>
\end{lstlisting}
\end{listing}

A \texttt{<page>} element corresponds to a single Wikidata page (e.g., an entity, a property, or a user talk) and contains a sequence of \texttt{<revision>} elements.
Each revision stores metadata (timestamp, user, comment) alongside a full snapshot of the page's state at that point in time (\texttt{<text>} tag). For item pages, the snapshot is structured following the JSON format described in~\cite{wikibase_json}. For entity pages, this includes labels, descriptions, and aliases in multiple languages, as well as statements with their rank, qualifiers, references, and sitelinks~\cite{wikibase_json}. 

\subsection{File Parsing}
\label{sec:tool-file-parsing}
As files store snapshots rather than deltas, extracting changes means comparing consecutive revisions of the same entity.
This process involves computing differences in nested JSON representations~\cite{wikibase_json}, managing five distinct data types and their associated metadata, and distinguishing changes at the levels of statement values, ranks, qualifiers, and references. 
\toolname{} focuses exclusively on Wikidata entities identified by QIDs; all other pages in the XML file are filtered out during parsing. Additionally, the tool filters the entities Wikidata has defined as Sandboxes, which allow users to try editing before committing changes to a real entity\footnote{\url{https://www.wikidata.org/wiki/Wikidata:Sandbox}}.

Given the scale of Wikidata -- individual XML files from the June 2025 dump can reach up to 6.3GB compressed and 77.85GB uncompressed -- we implemented a processing pipeline that incrementally reads XML files using the Python library lxml\footnote{\url{https://lxml.de/tutorial.html\#tree-iteration}}\footnote{\url{https://lxml.de/apidoc/lxml.etree.html\#lxml.etree.iterparse}}, rather than loading the full XML tree into memory. 
As shown in \autoref{fig:parser_arch}, the \textit{file parser} is in charge of decompressing and reading the XML files, where page elements are placed in a queue as they are read (\textit{page queue}) and consumed by worker processes that perform the change extraction (\textit{page parser}). Results are placed in an output queue (\textit{Output queue}) for batch insertion into the database by a writer process (\textit{DB writer}). This approach can be configured to process any number of files in parallel, with any number of worker processes per file, allowing the tool to scale to the available resources (see \autoref{sec:tool-config}).

\subsection{Change extraction}\label{sec:tool-change-extraction}

\autoref{lst:json_structure} shows how Wikidata models entity snapshots in their edit history dumps. Labels and descriptions are nested dictionaries containing one value per language, while statements (claims) are modeled as dictionaries where the key is the property ID and the value corresponds to the property values, each containing its own ID, rank, references, and qualifiers\cite{wikibase_json}.

To extract changes for a single entity, we compare consecutive revisions and extract changes to labels, descriptions, and a statement's value, rank, qualifiers, and references. 

For label and description changes, we extract the values for a single language (See \autoref{sec:tool-config} for the language configuration) from the previous and current revision and compare them; if they differ, a new change tuple is created with the respective edit type (CREATE, DELETE, or UPDATE).

\begin{listing}
\caption{Example of JSON structure for the entity Uruguay (Q77).}
\label{lst:json_structure}
\begin{lstlisting}[basicstyle=\ttfamily\scriptsize, language=XML]
{
  "id": "Q77",
  "type": "item",
  "labels": { "en": { "language": "en", "value": "Uruguay" } },
  "descriptions": {
    "en": { 
        "language": "en",  "value": "country in South America" 
    }
  },
  "aliases": {},
  "claims": {
    "P17": [
      {
        "id": "...",
        "mainsnak": {
            "snaktype": "value",
            "property": "P571",
            "datatype": "wikibase-item",
            "datavalue": {
                "value": {
                "time": "+2001-12-31T00:00:00Z",
                "precision": 11,
                "calendarmodel": "..."
            }
        },
        "type": "statement",
        "rank": "normal",
        "qualifiers": { "P580": [], },
        "references": [ { "snaks": [], } ]
      }
    ]
  }
}
\end{lstlisting}
\end{listing}

For statement changes, we extract all statements from both the previous and current revision, and partition them by property sets to identify: (1) added properties, (2) removed properties, and (3) properties present in both revisions.
For the added properties, we record all value, rank, qualifier, and reference edits as CREATE; for removed properties, we record all such edits as DELETE.

For properties appearing in both revisions, we compare values within each property via their identifier, yielding value-level deletions (present in previous but not current), value-level creations (present in current but not previous), and value-level updates (same value ID in both revisions). For each value present in both revisions, we compare the old and new values, and ranks; if any of these differ, we record the corresponding UPDATE operations.

For qualifiers and references within each statement, we use SHA-1 content-based hashes to identify their values and extract changes, since qualifier and reference values lack value identifiers compared to statement values (See \autoref{sec:change-model}).

For each \emph{qualifier property}, we hash each individual value (without data type metadata) and build a hash map keyed by the SHA-1 hash, with the corresponding value stored as the map entry.

For references, which can contain multiple statements, we compute a reference-level hash by hashing the set of all statements within that reference, ensuring that references with identical content produce the same hash. Then create a hash where the key is composed of (reference\_hash, property\_id, value\_hash) and the value is the reference statement value.

Set difference operations on these maps identify which qualifiers or references were added or deleted between revisions. 

This approach handles exact duplicates at both the qualifier and reference statement levels, as well as at the reference level.

\subsection{Storage Layer}\label{sec:tool-storage-layer}

As described in \autoref{sec:change-model}, for the different statement-edit granularities (value, qualifier, reference or rank), we have a fixed schema, determined by Wikidata's statement structure rather than anything that varies per entity or evolves over time.

Related work stores changes as change operations over triples~\cite{Zeginis2007,Zeginis2011}, as first-class citizens linked to the versions they apply to~\cite{RoussakisCSFS15}, as reified statements typed by a change ontology~\cite{PernelleSMT16}, as quads in a key-value store~\cite{Tanon2019}, or as sets of triple additions and deletions in JSON Lines~\cite{Wikidated2021}. The decisive difference is whether full graph versions must be materialized alongside the changes, which is impractical at Wikidata's scale.

We benchmarked the extended change cube in PostgreSQL and in QLever\cite{Qlever}
on the changes from 10 of the \numprint{2125} files in the June 2025 dump. QLever compressed indexes and vocabularies to 3.5GB, while PostgreSQL occupies 6.4GB. 
On the other hand, PostgreSQL loads and indexes in 155s, which is less than the 178s needed to serialize the tuples into RDF before QLever's own loading and indexing begins (171s), a factor of 2.25x overall. The same 25.5 million rows produce 195.1 million triples, all of which QLever requires materialized before indexing.
In summary, the strengths of RDF are schema flexibility and shared semantics, which is why Wikidata itself is published as an RDF graph. A change log derived from that graph has a different shape: its schema is fixed, and it grows append-only with each dump. We therefore store the log relationally.

The final database schema contains the following tables:
\begin{enumerate}
    \item Revision: stores revision metadata.
    \item Value change: stores edits (creates, deletes, and updates) to statements. 
    \item Rank change: stores edits (creates, deletes, and updates) to a statement's rank.
    \item Qualifier change: stores edits (creates and deletes) to qualifiers.
    \item Reference change: stores edits (creates and deletes) to references.
\end{enumerate}

A detailed description of our database schema is available in our code repository\footnote{\url{https://github.com/caroocortes/WiDiff}}.

\mypara{Example query}
To finalize this section, \autoref{lst:user_type_property} presents an example query that combines the \textit{value\_change} and \textit{revision} tables to analyze editing behavior of different users, for a specific property. The query returns the counts of creates, deletes, and updates for the different user types (bot, registered, or anonymous). 
\begin{lstlisting}[
style=sqlstyle,
language=SQL,
caption={SQL query combining \textit{value\_change} and \textit{revision} to retrieve counts of creates, deletes, and updates per user type (bot, registered, or anonymous) for a specific property}, label=lst:user_type_property, basicstyle=\footnotesize\ttfamily]
SELECT r.user_type, action, COUNT(*)
FROM 
    value_change vc JOIN revision r ON
    vc.revision_id = r.revision_id
WHERE 
    vc.property_id = NUMERIC_PID 
GROUP BY r.user_type, action
\end{lstlisting}

\subsection{Configuration}
\label{sec:tool-config}
\toolname is configured via a YAML file that controls three aspects of the extraction process.
\textit{Processing parameters} allow the user to specify the language for label and description change extraction, the number of files processed in parallel, and the number of worker processes per file. \textit{Entity filters} allow the user to exclude astronomical objects and scholarly articles from extraction, since these entities are predominantly added through automated imports and rarely edited afterward, making them less representative of the overall editing activity of Wikidata. 
The user can also exclude entities with fewer than a configurable number of value changes, further focusing the analysis on actively maintained entities. 

Finally, to optimize query performance, changes are stored in separate tables corresponding to each filter category (scholarly articles, astronomical objects, entities with fewer than \textit{X} value changes, and the remaining entities), rather than in a single table. This also enables users to focus their analysis on the different subsets of entities.

%% file: parts/6_change_data.tex
\section{Change Data}\label{sec:change_data}
We applied \toolname~to the June 2025 dump, which consists of \numprint{2125} compressed XML files totaling \numprint{2.2}~TB (compressed).
Extraction was performed with four files in parallel, each processed by two worker processes, and a database writer process per file (twelve processes in total), on a High-Performance Computing (HPC) cluster, utilizing nodes with AMD EPYC~7742 and Intel Xeon 8160/8352Y processors.
Even though XML files are streamed rather than loaded into memory, peak memory usage, for a single file, reached up to 181.5~GB across all parallel processes, since memory accumulates across the pipeline: pages waiting in the queue for worker processes to process them, extracted changes held in worker memory during processing, and results awaiting batch insertion in the output queues. Processing the entire revision history took 6.1 days of wall-clock time, without parallelization.
In total, we processed \numprint{120794819} entities and \numprint{2312348608} revisions, achieving an average throughput of 59~entities and \numprint{1126}~revisions per second, with an average runtime of 16~minutes per file. The average time to compute the diff between revisions was 0.37~milliseconds. All data extracted in our compressed representation amounts to 4.43~TB.

The reason for including filters for astronomical objects and scholarly articles was because these entity types exhibit editing patterns dominated by automated imports and statement creation with few updates. In particular, astronomical objects (8M entities) showed that 84\% of entities never received an update and 89\% of value changes are statement creations. On the other hand, scholarly articles amount to a total of 45M entities, with 56\% never updated, and 91\% of value changes being statement creations, with 70\% of revisions done by bots. Additionally, there are a total of \numprint{36137153} entities with fewer than ten value changes.

If one wants to focus only on entities receiving more than ten value changes that aren't scholarly articles nor astronomical objects (applying all filters described in \autoref{sec:tool-config}) the resulting dataset contains \numprint{31057469} entities, \numprint{743418656} revisions, \numprint{709506539} value changes, \numprint{616278002} rank changes, \numprint{1877386505} reference statement changes, \numprint{182486518} qualifier statement changes, and covers all entities' revisions from October 2012 to June 2025. Note that the number of revisions exceeds the number of extracted changes since we also store redirect revisions, which do not record any value change since the entity no longer has statements \wdlink{https://www.wikidata.org/w/index.php?title=Q2786851&oldid=2259027733}.

%% file: parts/7_change_analysis.tex
\section{Insights into Wikidata's Change Activity}\label{sec:analysis}

This section presents selected insights into Wikidata's change activity, by leveraging the data extracted with \toolname{} with the filters described in \autoref{sec:change_data}. 

\subsection{Distribution of Value Changes and Revisions}
\begin{figure}[hbtp]
    \centering
    \includegraphics[width=0.8\linewidth]{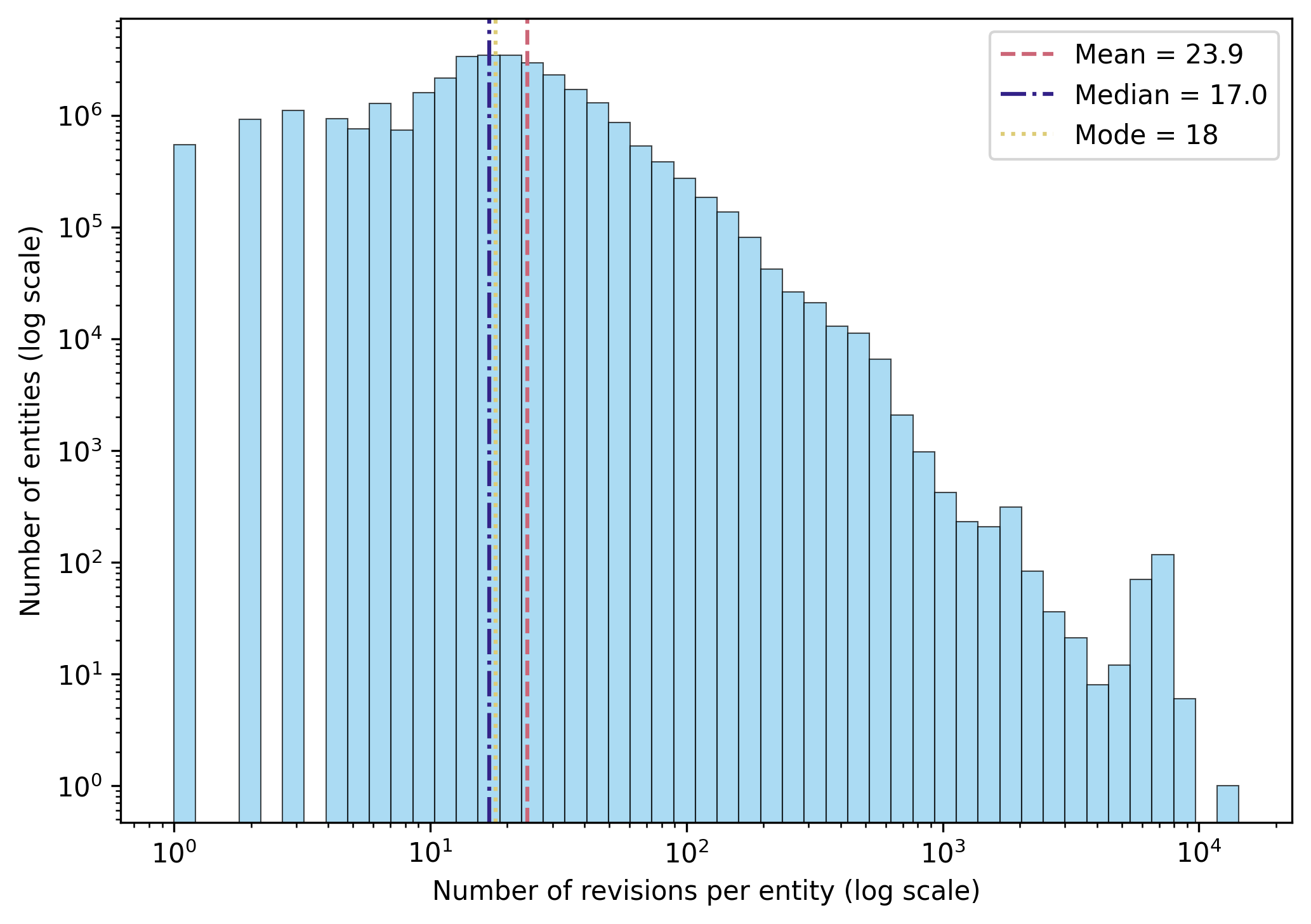}
    \caption{Distribution of revisions across entities.}
    \label{fig:revisions_distribution}
\end{figure}

\begin{figure}[hbtp]
    \centering
    \includegraphics[width=0.8\linewidth]{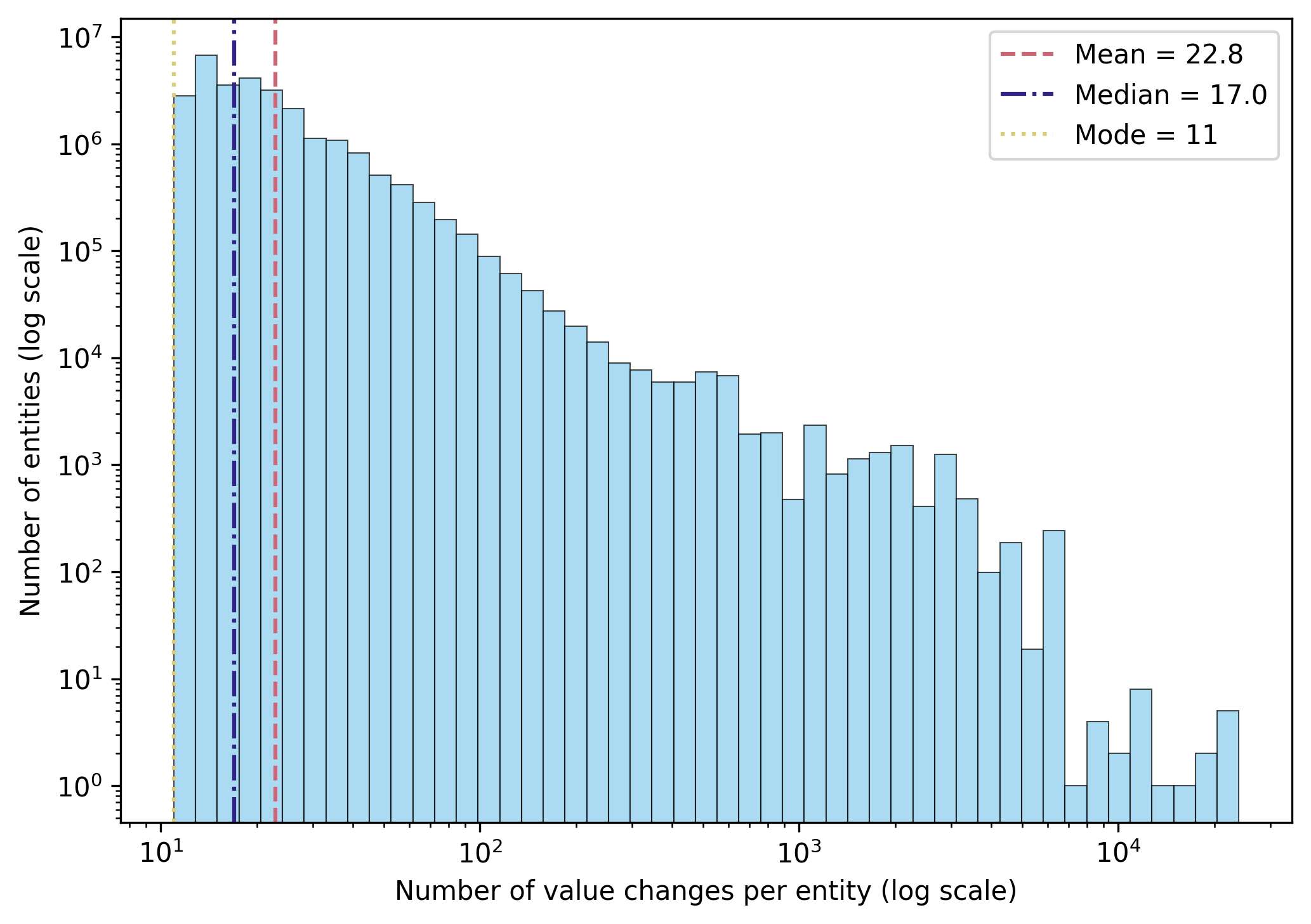}
    \caption{Distribution of value changes across entities.}
    \label{fig:value_changes_distribution}
\end{figure}

Unlike snapshot-based analysis, our change extraction lets us directly measure editing intensity across entities.
\autoref{fig:revisions_distribution} shows the distribution of revisions, while \autoref{fig:value_changes_distribution} shows the distribution of value changes (create, delete, and update) across all \numprint{31057469} entities. 
Notably, \numprint{546654} entities have only a single revision, which nonetheless contain an average of 13 value changes (create, delete, update), since a single revision can contain multiple edits simultaneously.
Overall, the average number of value changes per entity is 23, with an average of 24 revisions per entity. 
Moreover, \autoref{fig:value_changes_distribution} shows a right-skewed distribution, with 73\% of the entities having fewer than 23 value changes. On the other hand, for \autoref{fig:revisions_distribution}, 67\% of the entities have fewer than 24 revisions.

This distinction between revisions and underlying changes is not merely statistical; it can substantially affect how editing activity is interpreted. \entity{The Call of the Wild (Q476871)}, a novel by Jack London, has the highest revision count in our dataset (\numprint{14302}), yet it has only received changes to 61 properties. The property \schema{has version, edition or translation (P747)} has seen \numprint{22238} value changes over time -- illustrating that revision count alone substantially undercounts editing activity when revisions bundle multiple edits, a distinction only visible with change-level extraction.

Beyond edit volume, preserving full revision metadata -- including edit comments -- lets us identify likely automated imports. For instance, \entity{Africa (Q15)}, the entity with the earliest revision in our dataset (created October 29, 2012 at 17:03 UTC~\wdlink{https://www.wikidata.org/w/index.php?title=Q15\&oldid=16} by a registered user), is followed just hours later by \entity{Universe (Q1)}, created the same day at 18:18 UTC~\wdlink{https://www.wikidata.org/w/index.php?title=Q1\&oldid=103} with the revision comment ``Import'' -- suggesting an automated import rather than manual creation, and hinting that certain Q-identifiers may have been reserved in advance for bulk import.

\subsection{Deprecation and Promotion of statements' ranks}
Beyond aggregate volume statistics, our fine-grained change extraction enables provenance-style analysis, as shown next for rank deprecation and promotions.

We examined the deprecation of statements' ranks and found a lack of documented decisions, since only 35\% of rank deprecations have a qualifier with \schema{reason for deprecated rank (P2241)}. 
Among these, the most common values used for this qualifier are \schema{redirect (Q45403344)} (19\%) and \schema{withdrawn identifier value (Q21441764)} (15\%) -- both reflecting an identifier becoming outdated or superseded -- followed by \schema{link rot (Q1193907)} (9\%, URL no longer resolving), \schema{source known to be unreliable (Q22979588)} (5\%), and \schema{conflation (Q14946528)} (4\%, an erroneous merging of distinct concepts). 

Additionally, for statements with a preferred rank, only 4\% have a qualifier with \schema{reason for preferred rank (P7452)}, and the most common values for this qualifier are: \schema{most precise value (Q71536040)} (62\%), \schema{most recent value (Q71533355)} (22\%), and \schema{currently valid value (Q71536244)} (6\%).

Finally, only 0.07\% and 0.14\% of rank promotions and deprecations have been reverted\footnote{Let $c_{t_n}$denote a tuple in our data model representing a change that occurred at time $t_{_n}$ to a statement value or rank. A reverted edit $c_{t_1}$ occurs when there is a pair $(c_{t_1}, c_{t_2})$, with $t_1$ < $t_2$, and $c_{t_2}$ restores the value to the state before $c_{t_1}$ happened. Additionally, following the work of \cite{Nishioka2018, TanAIG14}, we require that this happens within four weeks}. This hints towards the reliability of these changes, since 71\% of rank deprecations are performed by registered users and 21\% by bots, while for rank promotions, 58\% is performed by bots and 40\% by registered users.

%% file: parts/8_conclusion.tex
\section{Conclusion}\label{sec:conclusion}
With this paper, we introduce \toolname, an open-source tool that extracts the complete edit history of Wikidata from historical dumps and stores it in a relational database, enabling a unified interface for large-scale analytical queries. Additionally, we made the dataset of extracted changes from the June 2025 dump available at \cite{full_data}.

Several directions remain for future work. 
First, the current tool does not extract changes to entity aliases or sitelinks, which we would like to incorporate in a future version. Moreover, the tool is currently only able to extract changes to labels and descriptions of a single language (see \autoref{sec:tool-config}); extending it to capture changes for labels and descriptions in multiple languages is a clear future improvement.
Second, while the current tool targets Wikidata, extending it to other openly available \kgs that expose change information, such as DBpedia, is a natural step toward a more comprehensive view of \KG evolution. 
However, this is not trivial as different \kgs expose their history in different formats. For instance, DBpedia extracts data from Wikipedia infoboxes and provides snapshots every four months, meaning changes that occur between snapshots are lost. 